\documentclass[aps,pre,preprint,showpacs,amssymb]{revtex4}
\usepackage{amsmath,graphics,epsfig}
\usepackage{pstricks}

\newcommand{\gae}{\hbox{\lower0.7ex\hbox{$\sim$}\llap{\raise0.4ex\hbox{$>$}}}}
\newcommand{\lae}{\hbox{\lower0.7ex\hbox{$\sim$}\llap{\raise0.4ex\hbox{$<$}}}}
\begin{document}
\title{Percolation transitions in two dimensions}
\author{Xiaomei Feng$^{1,2}$, Youjin Deng$^3$ and Henk W.J. Bl\"ote$^{1,4}$}
\affiliation{$^1$Faculty of Applied Sciences, Delft University of
Technology,\\  P.O. Box 5046, 2600 GA Delft, The Netherlands}
\affiliation{$^2$Nanjing University of Aeronautics and Astronautics 
29 Yudao St., 210016 Nanjing, P.R. China}
\affiliation{$^3$Physikalisches Institut, Universit\"at Heidelberg, 
Philosophenweg 12, 69120 Heidelberg, Germany}
\affiliation{$^4$Lorentz Institute, Leiden University,
  P.O. Box 9506, 2300 RA Leiden, The Netherlands}             
\date{\today} 
\begin{abstract}
We investigate bond- and site-percolation models on several two-dimensional
lattices numerically, by means of transfer-matrix calculations and Monte
Carlo simulations. The lattices include the square, triangular, honeycomb
kagome and diced lattices with nearest-neighbor bonds, and the square
lattice with nearest- and next-nearest-neighbor bonds. Results are 
presented for the bond-percolation thresholds of the kagome and diced
lattices, and the site-percolation thresholds of the square, honeycomb
and diced lattices. We also include the bond- and site-percolation
thresholds for the square lattice with nearest- and next-nearest-neighbor 
bonds.

We find that corrections to scaling behave according to the second
temperature dimension $X_{t2}=4$ predicted by the Coulomb gas theory
and the theory of conformal invariance. In several cases there is 
evidence for an additional term with the same exponent, but modified
by a logarithmic factor. Only for the site-percolation problem on the 
triangular lattice such a logarithmic term appears to be small or absent.
The amplitude of the power-law correction associated with $X_{t2}=4$
is found to be dependent on the orientation of the lattice with respect
to the cylindrical geometry of the finite systems.

\end{abstract}
\maketitle 

\section{introduction}
\label{Intro}
The bond-percolation model can be described by means of the partition sum
\begin{equation}
Z_{\rm bond}(p)=\prod_{\langle{ij}\rangle} \sum_{b_{ij}=0}^1
[(1-p) (1-b_{ij}) + p b_{ij}]=1
\label{zb}
\end{equation}
where the bond variables $b_{ij}$ are located on the edges of a lattice,
and labeled with the site numbers at both ends. The `bonds', i.e., the 
nonzero bond variables, form a network of which one may study the
percolation properties. Similarly, the site percolation problem
is described by
\begin{equation}
Z_{\rm site}(p)=\prod_{\langle{i}\rangle} \sum_{s_{i}=0}^1
[(1-p) (1-s_{i}) + p s_{i}]=1
\label{zs}
\end{equation}
In this case, the percolation problem is formed by adding `bonds'
between all pairs $(i,j)$ of neighboring sites that are both occupied 
($s_i=s_j=1$).

Eqs.~(\ref{zb}) and (\ref{zs}) specify that the bonds or
sites are occupied with independent probabilities $p$.
The values of these partition sums are trivial, but the percolation 
properties contained in these models are not. 
These properties, in particular for two-dimensional models, have been 
investigated by a considerable number of different approaches, see
\cite{MFS1,E,DV,HK,MNSS,BN,RMZ1,RMZ2,RMZ3,DB1} and references therein.
According to the universality hypothesis, some of these properties,
such as the critical exponents, are the same for different
two-dimensional lattices. Other properties, such as the percolation
threshold, are naturally dependent on the type of the lattice, as well
as on the number of neighbors to which a given site can form a bond. 
If not mentioned explicitly, we consider models with bonds between
nearest-neighbor sites only.

The present work reports some new findings, obtained by means of Monte
Carlo simulation and transfer-matrix methods. Monte Carlo simulation
was used in the cases of site percolation on the diced lattice,
and of bond percolation on the square lattice with nearest- and
next-nearest-neighbor bonds.

The outline of this paper is as follows. In Sec.~\ref{secnum} we sketch
our transfer-matrix and Monte Carlo methods. Section \ref{secres} 
describes the analyses and lists our results. The conclusions are
summarized and discussed in Sec.~\ref{seccon}.

\section{Numerical methods}
\label{secnum}
Our numerical analyses employ both transfer-matrix and Monte Carlo
techniques. Both approaches have their advantages. Transfer-matrix
calculation yield finite-size results of a high precision, typically
with error margins of the order of $10^{-12}$, which therefore allow 
the application of sensitive fitting procedures. However, the 
transfer-matrix results are restricted to rather small values of the 
finite-size parameter. In contrast, the Monte Carlo results can be
applied to much larger finite sizes, but they are also subject to
significant statistical errors. Which of the two techniques was applied
to specific models depended on considerations of effectiveness and
complexity. 
The site percolation problem on the diced lattice was investigated
by the Monte Carlo method, in view of the expected amount of work 
involved in writing the various sparse-matrix multiplication subprograms 
in a transfer-matrix algorithm for the diced lattice, which has
inequivalent sites. While the transfer-matrix method usually yields
relatively accurate results, this appeared not to be the case for the
bond-percolation problem on the square lattice with crossing bonds.
For this reason we also investigated this problem using the Monte
Carlo method.

\subsection{The transfer matrix}
\label{sectm}
The present transfer-matrix calculations apply to models wrapped on an
infinitely long cylinder, with a finite circumference $L$. 
For the bond-percolation model, which can be considered as the special
case $q=1$ of the random-cluster representation of the $q$-state Potts 
model \cite{KF}, we may conveniently use the numerical methods
developed earlier for the random-cluster model.
Our basic approach \cite{BND} is close in spirit to that of Derrida and
Vannimenus \cite{DV} for the percolation model.
Ref.~\onlinecite{BNi82} describes in detail how the state of
connectivity of the sites on the end row of the cylinder can be coded by
means of an integer (1, 2, 3, $\cdots$) that serves as a transfer-matrix
index. That work also describes the sparse-matrix decomposition of the
transfer matrix. However, the various lattice structures investigated
here require modifications of the sparse-matrix methods described there.
In addition, the model with nearest- and next-nearest-neighbor
bonds violates the condition of `well-nestedness' described in
Ref.~\onlinecite{BNi82}, so that new coding and decoding algorithms had 
to be devised.
The transfer matrices for the site-percolation models require
different modifications that, again, depend on the lattice structure. 
It is, in general, necessary to store the occupation number (0 or 1)
of the sites on the end-most row of the lattice, as well as the state
of connectivity of the occupied sites. This combined information can
also be coded as an integer that plays the role of a transfer-matrix
index, using the methods described in Ref.~\onlinecite{QDB}.
However, the state of connectedness of the occupied sites on the last
row may, depending on the lattice geometry, be subject to an additional
condition. If the sites on this row are nearest neighbors, such as
for the square lattice with the transfer direction along one set of
lattice edges, then two adjacent, occupied sites must belong to the
same cluster. This reduces the number of possible connectivities.
To take advantage of this reduction, the coding-decoding algorithms
for the square-lattice model were modified. For the square-lattice
site-percolation model with transfer in the diagonal direction,
the situation is different, and the algorithms of Ref.~\onlinecite{QDB}
had to be used. To save memory and computer time, sparse matrix
decompositions were applied in all cases.  A full description of all
these algorithms is beyond the scope of this paper; we trust that
it is sufficient to mention that all further relevant details are
contained in or follow from the explanations given here and in
Refs.~\onlinecite{BNi82,QDB,BWW,KBN,TI}.

The connectivities used here include those of the `magnetic' type,
i.e., they carry the information which of the sites of the  end row
are connected by a percolating path to a far-away site, say on the first
row of the lattice.

Using a  computer with 8 gigabyte of fast memory, our algorithms can
perform transfer matrix calculations in connectivity spaces of linear 
dimensions up to the order $10^{8}$. Some details concerning the 
largest system sizes that could thus be handled, and the corresponding
transfer-matrix sizes, appear in Tab.~\ref{tab:tmdat}.

\begin{table}
\caption{Some details about the transfer-matrix calculations on the
various models. Transfer directions are given with respect to a set of
lattice edges. Included are the largest system sizes and the linear
size of the transfer matrix for that system, as well as the corresponding
size of the largest sparse matrix. The entry `8-nb square' refers to the
square lattice with nearest- and next-nearest-neighbor bonds.}
\label{tab:tmdat}
\begin{center}
\begin{tabular}{|l|c|l|c|r|r|}
\hline
Lattice    & type & direction&  $L_{\rm max}$ & max size & sparse size \\
\hline
kagome     & bond & perpendicular  &    13    &  5943200 &   22732740  \\
square     & bond & parallel       &    15    & 87253605 &   87253605  \\
square     & bond & diagonal       &    14    & 22732740 &   87253605  \\
triangular & bond & perpendicular  &    14    & 22732740 &   87253605  \\
8-nb square& bond & parallel       &    10    &   678570 &   27644437  \\
square     & site & parallel       &    16    &  6903561 &   57225573  \\
square     & site & diagonal       &    12    & 26423275 &  125481607  \\
honeycomb  & site & parallel       &    12    & 26423275 &  125481607  \\
triangular & site & perpendicular  &    17    & 19848489 &   57225573  \\
\hline
\end{tabular}
\end{center}
\end{table}

The eigenvalue problem of the transfer matrix reduces in effect to 
separate calculations in the magnetic and nonmagnetic
sectors. The calculation of the largest eigenvalue in the nonmagnetic
sector trivially yields $\Lambda_0=1$.
The transfer-matrix construction enables the numerical calculation of
the magnetic eigenvalue $\Lambda_1$ as described earlier \cite{BNi82}.

The analysis of these magnetic eigenvalues uses 
Cardy's mapping \cite{JCxi} which establishes an asymptotic relation
between the magnetic eigenvalue and the exact magnetic scaling
dimension $X_h$. Furthermore we employ knowledge of the exact
critical exponents from the Coulomb gas theory \cite{BN} and the 
theory of conformal invariance \cite{JC}. These results establish that
$X_h=5/48$ for percolation models, and that the first and second
thermal dimensions are equal to $X_t=5/4$ and $X_{t2}=4$ respectively.

\subsection{Monte Carlo calculations}
\label{secmc}
\begin{figure}
\begin{center}
\leavevmode
\epsfxsize 12.6cm
\epsfbox{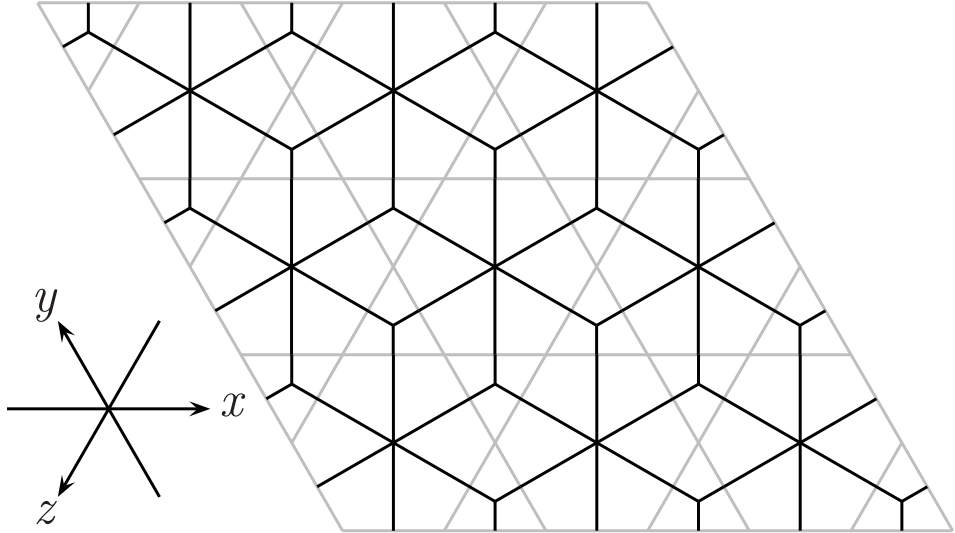}
\end{center}
\caption{The diced lattice (full lines) and its dual, the kagome lattice
(thin lines). The three main axes are labeled $x$, $y$ and $z$. }
\label{fig1}
\end{figure}

We employed Monte Carlo simulations for the site-percolation problem
on the diced lattice, and for the bond-percolation model on the square
lattice with nearest- and next-nearest-neighbor bonds.
The finite systems were defined in an $L \times L $ periodic geometry,
in the case of the diced lattice on the basis of a rhombus with an
angle $2 \pi/3$ between the main axes, as illustrated in Fig.~\ref{fig1}.
The system, including its periodic structure,
displays a hexagonal symmetry. Thus, for the definition of the periodic
box one has in fact the freedom to choose any two out of three main axes
separated by angles $2 \pi/3$.
For the square lattice we employed a 
periodic box with the usual square symmetry, with only two main axes.
A Metropolis-like procedure was applied:
one visits the sites or bonds sequentially, and randomly decides with
probability $p$ whether it is occupied; clusters are then constructed
on the basis of these occupied site or bond variables. We  employed a
random generator based on binary shift registers. To avoid errors
resulting from the use of single short shift registers \cite{disp},
we used the modulo-2 addition of two independent shift registers
with lengths chosen as the Mersenne exponents 127 and 9689. This type
of random generator is well tested \cite{SB}.

For a sufficiently long series of percolation configurations thus
obtained, we sampled the wrapping probability $P$ that a configuration
has {\it at least} one cluster that wraps across a periodic boundary
and connects to itself along {\it any} of the aforementioned main axes.
This is done for a range of values $p$ of the site- or bond probabilities.

For the analysis of the data for the model on the diced lattice, it is
helpful that the value of the wrapping probability $P_c$ is exactly
known for the periodic rhombus geometry of the critical triangular
bond percolation model as $P_c=0.683946586 \cdots $~\cite{Ziff-99}
(note that $\pi_{+}$ in the latter paper represents our $P_c$).
It applies in the limit of large system size $L$ and is believed
to be universal, i.e., it also applies to the diced lattice which also
has a hexagonal symmetry.  For the periodic square geometry, 
the universal value $P_c$ is exactly known as
$P_c=0.690473725 \cdots $~\cite{HTPinson,Ziff-99}.

The simulations were performed for $15$ system sizes in the range
$ 4 \leq L \leq 256$;
about $21 \times 10^9$ samples were taken for each $L$ for $L \leq 64$,
and $6 \times 10^9$ samples for $L=128$ and 256.

\section{Results}
\label{secres}
The analysis of the  numerical finite-size data was done by means
of well-documented finite-size scaling methods \cite{FSS}. 
We describe the procedures followed for the transfer-matrix and Monte
Carlo data separately.

\subsection{Percolation thresholds}
\subsubsection{Transfer matrix results}
\label{pthres}
The data analysis was performed on the basis of the scaled gap 
\begin{equation}
X_h(p,L) \equiv \frac{\zeta L \ln(\Lambda_0/\Lambda_1)}{2 \pi} 
\end{equation}
where $\zeta$ is the geometric factor defined as the ratio between 
the lattice unit in which the finite size $L$ is expressed, and the
thickness of the layer added to the lattice by a transfer-matrix
operation. According to finite-size scaling, the scaled gap behaves,
near the percolation threshold $p_c$, as
\begin{equation}
X_h(p,L) = X_h +a (p-p_c) L^{2-X_t} + b L^{2-X_{t2}} + \cdots 
\label{Xsc}
\end{equation}
where $a$ and $b$ are model-dependent parameters.
It follows from the definition of $p_c(L)$ as the solution of the equation
\begin{equation}
X_h(p_c(L),L) = X_h
\end{equation}
and from Eq.~(\ref{Xsc}) that
\begin{equation}
p_c(L) \simeq p_c + c L^{X_t-X_{t2}}
\end{equation}
with $c=-b/a$. Since $X_t-X_{t2}=-11/4$, the finite-size estimates
$p_c(L)$ should converge rapidly to $p_c$. In fact, the numerical data
allow independent fitting of the exponent $X_t-X_{t2}$ and thus provide
an independent confirmation of its value $-11/4$. On this basis one can,
for instance, rule out a leading correction with exponent $-7/4$, such
as would be generated by a hypothetical integer dimension $X=3$. Assuming
$X_{t2}=4$, improved convergence of the $X_h$ estimates is obtained by
iterated power-law fitting as described in Ref.~\onlinecite{BNi82}.
After a first fitting step with exponent $-11/4$, the next iteration
step yielded, in most cases, an exponent with approximately the same value,
which suggests that Eq.~(\ref{Xsc}) should be replaced by
\begin{equation}
X_h(p,L) = X_h +a (p-p_c) L^{2-X_t} + (b+d \ln L ) L^{2-X_{t2}} + \cdots
\label{Xsc1}
\end{equation}
The appearance of such logarithmic terms is consistent with renormalization
theory for scaling relations involving integer exponents \cite{Wegner}.
Final estimates for the percolation threshold were obtained from another
power-law iteration step.
These results are shown in Tab.~\ref{tab:tmres}, together with error
estimates in the last decimal place.

These error estimation of the extrapolated results requires considerable
attention. While subsequent iteration steps eliminate successive
corrections, the remaining corrections are, in principle, unknown.
Fortunately, the apparent convergence of the fits indicates that they
decay rapidly, i.e., with rather large and negative exponents of $L$.
The error estimates can be based on the differences between the results 
from the last iteration step for a few of the largest available system
sizes. The rapid decrease of these differences with increasing $L$
suggests that the error of the extrapolated result is of the same order
as the differences for the largest $L$ values. However, it is obvious
that a single estimate of this type is not very reliable, and
one should search for additional evidence. First, one can vary the
fitting procedure, for instance one can fix the correction exponent
at $-11/4$ in the second or the third iteration step, or treat it as
a free parameter. Another variation is to use, in the second iteration
step, a fit of the form given by Eq.~(\ref{Xsc1}). These procedures
yielded consistent results, and also provide independent data on the
accuracy of the extrapolations. Furthermore, the amplitude of the
correction term as evaluated in the last iteration step should behave
sufficiently regularly as a function of $L$. If not, the differences
in the last iteration step are not a reliable basis for the error
estimation. When these conditions were satisfied, we took the error
estimate equal to a few times the typical difference between the
results for the largest two systems.
To provide some actual information about the apparent convergence of
the percolation thresholds, we list the largest-$L$ differences of the 
finite-size estimates of the original data, and of the first, second,
and third iteration steps, for the case of the site-percolation problem
on the square lattice, with transfer parallel to the edges. While these
differences depend on the fit procedure, they typically amount to
$2\times 10^{-5}$, $10^{-6}$, $10^{-7}$, and less than $10^{-8}$
respectively. Furthermore, the amplitude of the last power-law step
appears to tend smoothly to a constant and gives no sign of, for
instance, an extremum as a function of $L$. 

\begin{table}
\caption{Summary of percolation thresholds of some two-dimensional  
 lattices. The symbol $z$ is the coordination
 number; $p_c^{\rm bond}$ and $p_c^{\rm site}$ represent the critical 
 bond- and site-occupation probabilities, respectively. Errors in the
 last decimal place are given in parentheses. The value 0 is given
 in those cases where the percolation threshold is exactly 
 known \cite{E}. The remaining entries were obtained from the literature
 as indicated by the reference listed, or by the present numerical
 analyses, which use  Monte Carlo simulations (as indicated by MC
 in the source column) or transfer-matrix calculations (as indicated by
 TM). The bond percolation threshold for the diced lattice follows from 
 a duality transformation of the kagome lattice model, and did therefore
 not require separate
 calculations. Similarly, the entry for the site-percolation threshold
 for the eight-neighbor square lattice follows from that for the matching
 lattice, i.e., the entry for the four-neighbor model.}
\label{tab:tmres}
\begin{center}
\begin{tabular}{|l|c||l|c||l|c|}
\hline
Lattice   &$z$&$p_c^{\rm bond}$&  source  & $p_c^{\rm site}$&source       \\
\hline
triangular& 6 &0.3472964... (0)&  exact   &0.5          (0) &exact        \\
\hline
honeycomb & 3 &0.6527036... (0)&  exact   &0.6970402    (1) &  TM         \\
\cline{5-6}
          &   &                &          &0.697043     (3) &\cite{RMZ1}  \\
\cline{5-6}
          &   &                &          &0.69704024   (4) &\cite{WZhang}\\
\hline
kagome    & 4 &0.52440499 (2)  &  TM      &0.6527036... (0) & exact       \\
\cline{3-4}
          &   &0.5244053  (3)  & \cite{RMZ1}  &             &             \\
\cline{3-4}
          &   &0.52440503 (5)  &  MC          &             &             \\
\hline
diced &~~3,6~~&0.47559501 (2)  &  TM,d    &0.58504627   (6) & MC          \\
\hline
square    & 4 &0.5        (0)  & exact    &0.59274605  (3)  & TM          \\
\cline{5-6}
          &   &                &          &0.59274603  (9)  &\cite{ML}    \\
\cline{5-6}
          &   &                &          &0.59274606  (5)  & MC          \\
\hline
square    & 8 &0.250369   (3)  &  TM      &0.40725395  (3)  & TM,m        \\
\cline{3-4}
          &   &0.25036834 (6)  &  MC      &                 &             \\
\hline
\end{tabular}
\end{center}
\end{table}

\subsubsection{Monte Carlo results}
The numerical results for the wrapping probability $P$ defined in
Sec.~\ref{secmc} were fitted, using the least-squares criterion, by
means of the finite-size-scaling formula
\begin{equation}
 P (p,L)=P_c +a_1 (p-p_c) L^{y_t}
+ a_2 (p-p_c)^2 L^{2y_t} + b_1 L^{y_i} + b_2  L^{y_i-1}
+ c (p-p_c) L^{y_t+y_i} \; ,
\label{fit_q}
\end{equation}
where $y_t=2-X_t=3/4$ is the temperature exponent,
and $y_i=2-X_{t2}=-2$ is the irrelevant exponent \cite{BN} describing
the corrections to scaling.

We simulated the bond-percolation model on the triangular lattice
right at the exactly known critical point
for $15$ values of $L$ in range $4 \leq L \leq 512$. The number of 
samples is about $1 \times  10^{10}$ for system sizes $L \leq 256$, 
and $2 \times 10^9$ for $L=512$. The wrapping probability $P$ was 
fitted by Eq.~(\ref{fit_q}), excluding the $p$-dependent terms. 
We obtained $P_c=0.683947$ $(3)$, in good agreement with the exact
result $0.683946586 \cdots $~\cite{Ziff-99}.

For the diced lattice, the asymptotic critical wrapping probability
was fixed at the exact value.
The $P(p,L)$ values appear to be well described by
Eq.~(\ref{fit_q}) for system sizes not smaller than the
minimum size $L_{\rm min} = 16 $. Satisfactory fits (as judged from
the $\chi^2$ criterion) could also be obtained for smaller values of
$L_{\rm min} = 16 $ when additional corrections were included with
exponents $y_i-2$ and $y_i-3$. These fits are quite stable with respect
to variation of $L_{\rm min}$,
and yield the site-percolation threshold of the diced lattice
as $p_c= 0.58504627$ $(6)$. 

Also for the bond percolation on the square lattice with nearest- and
next-nearest-neighbor bonds we fitted the $P(p,L)$
data by Eq.~(\ref{fit_q}), but with the wrapping probability fixed
at $P_c=0.690473725 \cdots $~\cite{HTPinson,Ziff-99}.
Satisfactory fits were obtained for $L \geq 6 $,
and yield the bond-percolation threshold as $p_c =0.25036834$ $(6)$.
The estimates for $p_c$ are included in Tab.~\ref{tab:tmres}.

The estimation of the uncertainty margin in $p_c$ is relatively
straightforward. The  Monte Carlo runs are divided in 2000 subruns,
and the error in the average of a run follows from the standard deviation
of the subrun averages. The multivariate analysis that determines
$p_c$ thus also produces the statistical error in this quantity.
However, the actual error is still subject to the effects of correction
terms not included in Eq,~(\ref{fit_q}).

Such additional correction terms decay rapidly with the system size, 
so that the finite-size cutoff parameter $L_{\rm min}$ is reasonably
well determined by the $L_{\rm min}$-dependence of the residual
$\chi^2$ of the fits. This finite-size cutoff parameter naturally
depends on the number of finite-size corrections included.
Thus many fits were made to determine each $p_c$, varying the
number of correction terms in the fit formula, and varying the
minimum system size below which the finite-size data were excluded.
he errors quoted are such that the margins include
all one-standard-deviation lower and upper bounds of several different
fits, using different fit formulas as well as a range of different
accetable  values of $L_{\rm min}$ for each fit formula.

\subsection{Corrections to scaling}
The analysis of the finite-size data to determine the percolation
thresholds in Sec.~\ref{pthres} indicated that there are corrections
described by an irrelevant scaling dimension $X_{t2}$ close to the value
$4$ predicted by the Coulomb gas analysis \cite{BN} and the Kac
formula \cite{AAB,FQS,Kac}. However, the analysis also suggested that
corrections governed by this exponent contain a logarithmic correction
factor.
The models for which the percolation threshold is exactly known,
such as the bond-percolation model on the square lattice and the
site-percolation model on the triangular lattice, allow a study of
the finite-size dependence of the scaled gap $X(p_c,L)$ at the exact
critical point. In that case the corrections are due only to the 
irrelevant fields, and additional errors due to the uncertainty
of the percolation threshold are eliminated. Analysis of the scaled 
gap will purportedly reveal the nature of the corrections associated
with the leading irrelevant exponent.
In order to focus more precisely on possible logarithmic terms, we
defined the model-dependent quantity $C(L)$ as
\begin{equation}
C(L) \equiv (X_h(p_c,L) -X_h) L^{2}
\label{cora}
\end{equation}
which serves as an estimate of the amplitude of the finite-size correction
term in $X_h(p_c,L)$ if a logarithmic term is absent.
For a few models with exactly known percolation thresholds, we calculated
finite-size data for $C(L)$ and applied a fit of the form
\begin{equation}
C(L) \approx C + A \ln L +B L^{-r}
\label{coraf}
\end{equation}
First the parameters $C$ and $A$ were solved from two consecutive values
of $C(L)$, with the amplitude $B$ set to zero. The third term, which is
treated as a perturbation, is then taken into account in the second step
by means of an iterated power law fit as described in Ref.~\onlinecite{BNi82}.
This approach led to a series of apparently well-convergent estimates of the
constant $C$ and the amplitude $A$. These are shown in Tab.~\ref{tab:ab}.

This analysis was unable to yield good estimates of the exponent $r$,
which indicates that there exist further correction terms,
in addition to those listed in Eq.~(\ref{coraf}). However, the data
were insufficient to obtain more quantitative information.
\begin{table}
\caption{Results of the analysis of the corrections to scaling in the
quantity $X_h(L)$ for a few exactly solved bond- or site-percolation
models. Transfer directions 
are given with respect to a set of lattice edges and specify the 
orientation of the lattice with respect to the axis of the cylinder on
which the model is wrapped. Results are shown for the amplitudes $C$ and
$A$ of the $L^{-2}$ and the $L^{-2}\ln L$ terms respectively.}
\label{tab:ab}
\begin{center}
\begin{tabular}{|l|c|l|r|r|}
\hline
Lattice    & type & direction      &   $C$~~~~~ &       $A$~~~~~    \\
\hline
square     & bond & parallel       &   0.0306  (1) & $-0.0054$ (1)  \\
square     & bond & diagonal       & $-0.0205$ (1) & $-0.0027$ (1)  \\
triangular & bond & perpendicular  & $-0.0037$ (1) & $-0.0036$ (1)  \\
triangular & site & perpendicular  &   0.0195  (1) &   0.0000  (1)  \\
\hline
\end{tabular}
\end{center}
\end{table}
The difference between the two entries for the amplitude $A$ for the
square lattice is, at least approximately, equal to 2. This factor
may be attributed to the difference of a factor $\sqrt 2$ in the 
length units of the finite size $L$ for the two cases (an edge or a
diagonal of the square lattice). This would suggest that the amplitude
$A$ of the logarithmic term is, unlike the amplitude $C$, independent
on the orientation of the finite direction of the square lattice
in the cylindrical geometry.

\section{Conclusion}
\label{seccon}
We obtained new results for the percolation thresholds of several
two-dimensional models. The results are, as far as they overlap
with the literature, generally consistent with existing results,
and the error margins are somewhat reduced. Although our result for
the bond-percolation threshold of the kagome lattice model lies
remarkably close to an approximate result given by Scullard and
Ziff \cite{RMZ3}, the difference is quite significant, in agreement
with the conclusions of these authors \cite{RMZ3}.

In our numerical analysis of the transfer-matrix data, we made use
of the universality hypothesis, i.e., in all cases we assumed the
validity  of the exact results for the scaling dimensions of  the
percolation models in two dimensions \cite{BN}. However, we are
able to support this assumption considerably. In addition to
Eq.~(\ref{Xsc}), we may use `phenomenological renormalization'
\cite{MPN}  to determine the critical points, so that we no longer
make use of prior knowledge $X_h=5/48$. This approach yields the
same critical points, within error margins that are a few times
larger than those listed in Tab.~\ref{tab:tmres}. The estimates
of the scaled gaps produced by the phenomenological renormalization
approach match the value 5/48 up to several decimal places, and give
no sign of deviations from universality.

The analysis of the corrections to scaling is in agreement with the
irrelevant scaling dimension $X_{t2}=4$, but showed  the existence of
a contribution with a logarithmic factor in the transfer-matrix data
for the scaled magnetic gap. The amplitude of this contribution is
strongly model-dependent, and possibly vanishes for the triangular
site-percolation model. This raises the question in what sense the
latter model could be special. It may be argued that it is the model
with the highest symmetry investigated here; it has a $\pi/3$ rotational
symmetry as well as a form of self-dual symmetry because of the
matching-lattice argument \cite{E}. A corresponding logarithmic term
could possibly also be present in the quantity $P(p,L)$, but we were
unable to confirm its existence from our Monte Carlo data.

Furthermore, we recall that the bond-percolation model with crossing
bonds (8 neighbors) lives in an extended space of connectivities because
the condition of well-nestedness \cite{BNi82} no longer applies.
Accordingly one may postulate that these non-well-nested connectivities 
introduce another irrelevant field, and that additional corrections 
described by a new scaling dimension would appear.
However, we did not find any clear sign of such new corrections.

Finally we remark that the present logarithmic terms are unrelated
to those reported by Adler and Privman \cite{AP}, which apply to 
some leading singularities. Logarithmic factors naturally appear in
quantities involving the mean cluster size.  The free energy of the
random cluster model also serves as the generating function of
percolation properties. The mean cluster size can be obtained
by differentiation of the random-cluster partition sum or the free
energy to the number of Potts states $q$. The $q$-dependence of the
critical exponents then provides a mechanism that introduces such
logarithmic factors in some critical singularities.

\acknowledgments
We acknowledge discussions with Profs. J. C. Cardy and R. M. Ziff.
YD thanks for the support  of the Alexander von Humboldt Foundation
(Germany). HB thanks the Lorentz Fund (The Netherlands) for support.

\end{document}